\documentclass[reprint, aps, pra]{revtex4-1}
\usepackage[margin=1in]{geometry}                  
\usepackage{amsmath}
\usepackage{natbib}
\usepackage{graphicx}
\usepackage{epstopdf}
\usepackage{amssymb}
\usepackage{url}
\usepackage{bm}
\usepackage{subfigure}
\usepackage{dcolumn}
\usepackage[hidelinks]{hyperref}
\usepackage{mathrsfs}
\begin{document}
\title{Simulation of Anderson localization in two-dimensional ultracold gases for point-like disorder}
\author{W. Morong and B. DeMarco}
\affiliation{University of Illinois at Urbana-Champaign, Dept. of Physics, 1110 W Green St, Urbana, IL 61801}
\date{\today}                                           

\begin{abstract}
Anderson localization has been observed for a variety of media, including ultracold atomic gases with speckle disorder in one and three dimensions. However, observation of Anderson localization in a two-dimensional geometry for ultracold gases has been elusive.  We show that a cause of this difficulty is the relatively high percolation threshold of a speckle potential in two dimensions, resulting in strong classical localization.  We propose a realistic point-like disorder potential that circumvents this percolation limit with localization lengths that are experimentally observable.   The percolation threshold is evaluated for experimentally realistic parameters, and a regime of negligible classical trapping is identified.  Localization lengths are determined via scaling theory, using both exact scattering cross sections and the Born approximation, and by direct simulation of the time-dependent Schr\"{o}dinger equation.  We show that the Born approximation can underestimate the localization length by four orders of magnitude at low energies, while exact cross sections and scaling theory provide an upper bound.  Achievable experimental parameters for observing localization in this system are proposed.
\end{abstract}

\maketitle

\section{Introduction}

Anderson localization (AL) \cite{Anderson1958, Kramer1993}---the preclusion of wave propagation in a disordered medium by interference---has been observed in many settings, including, e.g., for light and sound \cite{Schwartz2007, Hu2008}. Recent experiments observing AL for ultracold atomic gases expanding in disordered optical speckle potentials show promise for gaining enhanced understanding of how microscopic disorder characteristics affect localization and the interplay with inter-particle interactions. In particular, these systems allow for independent control of inter-atomic interactions and the disorder strength and correlation length. Localization has been observed and its dependance on these disorder parameters studied for gases confined to one-dimensional geometries \cite{Billy2008, Roati2008} and in three dimensions \cite{Kondov2011, Jendrzejewski2012,PhysRevLett.111.145303}. In two-dimensional gases, however, the classical diffusive regime \cite{Robert-de-Saint-Vincent2010} and the impact of disorder on superfluids have been explored \cite{Krinner2012, Beeler2012}, but AL has not yet been observed.  Studying localization in two dimensions using ultracold gases is especially desirable given the many outstanding questions regarding localization in two-dimensional electronic solids \cite{kravchenko2010,goldman2010,RevModPhys.73.251}.

In this paper, we discuss how classical trapping effects complicate the observation of AL for speckle disorder in two dimensions.  We demonstrate how point-like disorder [Fig. 1] avoids these problems. Through a combination of analytical and numerical simulation we show that a disordered potential of this type would enable experimental observation of two-dimensional AL for ultracold gases. Point-like, two-dimensional disorder and time-dependent simulations of localization have not been considered in previous theoretical studies of ultracold systems \cite{Kuhn2007, Shapiro2012, Piraud2013}. Like previous studies, we consider an ideal two-dimensional geometry and ignore inter-particle interactions and quantum statistics.  As our focus is on realistic experimental conditions, we have limited our investigation to experimentally accessible disorder strengths and energy scales for ultracold, spin-polarized $^{40}$K atoms. We expect that this technique can be straightforwardly extended to other experimental configurations.

\begin{figure}[h!]
\includegraphics[width=0.5 \textwidth]{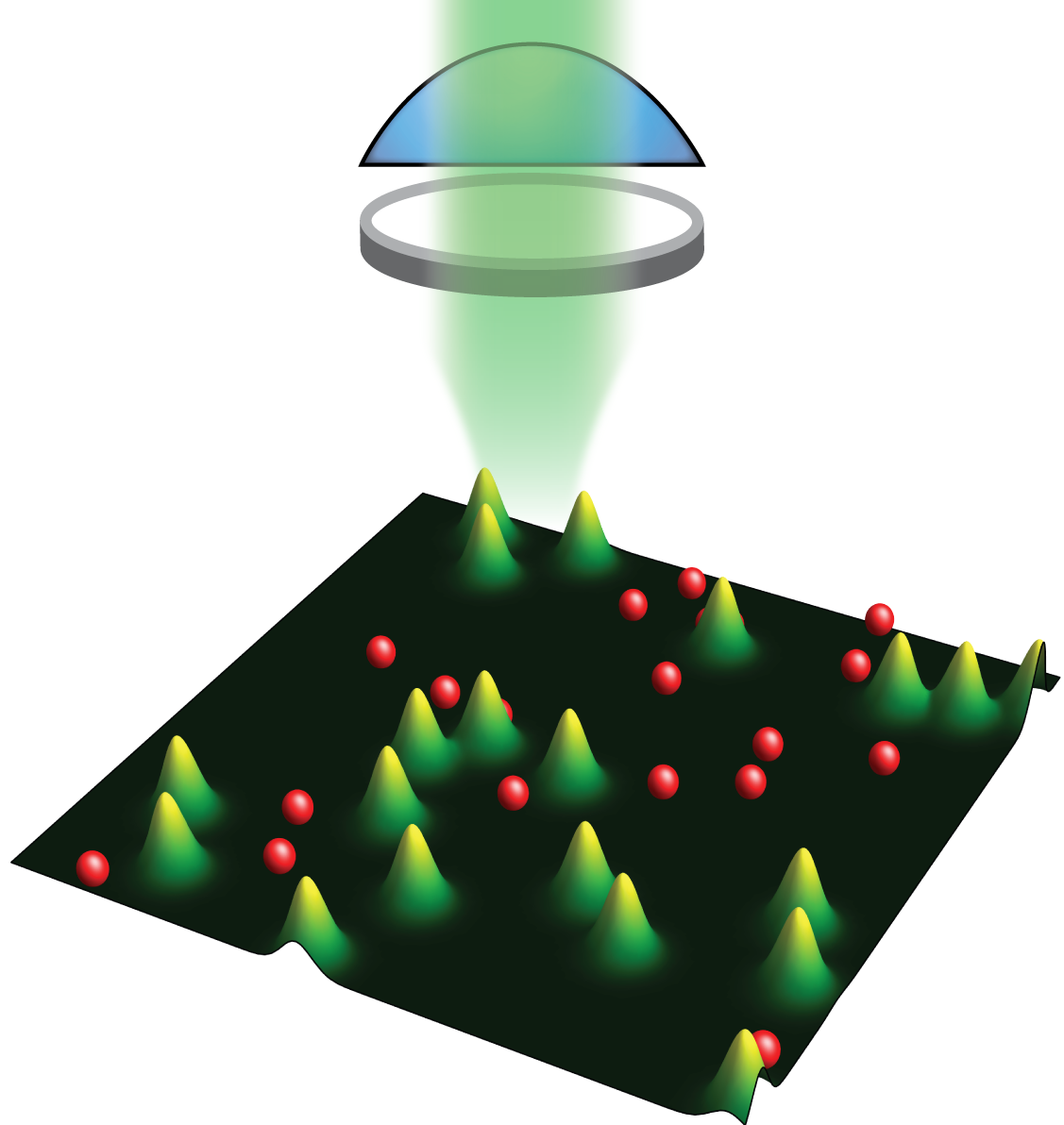}
\caption{Schematic representation of experimental implementation of point-like disorder. Ultracold atoms are confined to a quasi two-dimensional geometry using a sheet of far-detuned light (not shown). The disordered potential is generated by an additional laser beam (green) that passes through a holographic optic (disc) and is focused on the atoms (red spheres).  The atoms experience a disorder potential consisting of a random arrangement of Gaussian barriers.}
\label{setup}
\end{figure}

This paper is organized as follows: in Sec. II we describe classical trapping in two-dimensional speckle potentials using percolation theory, and we show how point-like disorder can avoid this problem for realistic experimental parameters. In Sec. III we identify experimentally accessible parameters for localization in point-like disorder using scaling theory and the Born approximation, which we show may fail in this regime.  Therefore, in Sec. IV, we calculate exact differential cross sections using numerical simulations to more accurately determine localization lengths via scaling theory.  We also use exact time-dependent simulations of wavepacket propagation in point-like disorder to determine localization lengths as would be observed in an experiment that allows a gas to expand into a disordered potential.

\section{Percolation in Speckle and Point-like Disorder}

In this section, we examine the classical trapping properties of speckle and point-like disorder potentials in two dimensions using percolation theory, which is the study of the random growth of interconnected regions in networks. These regions grow in size as a parameter is varied until a transition at the percolation threshold, at which a connected region spanning the system is formed \cite{Essam1980, Mendelson1997}. In the context of atom transport in a two-dimensional disordered potential, the problem is easily visualized in this way: the disorder potential forms a potential landscape that constrains the atomic motion.  Atoms with kinetic energy higher than the percolation threshold will travel freely through the potential, but below a critical fraction of the average potential energy atoms will be trapped in potential minima of finite size \cite{Zallen1971}. If this percolation threshold occurs near the same energy scale as Anderson localization, AL may not be detectable or experimentally distinguishable from classical trapping.

Because of their ease of creation and simple statistical properties, optical speckle fields \cite{Goodman2007} have been used to generate disorder in virtually all ultracold atom experiments exploring disorder-induced effects, see, e.g., Refs. \citenum{Billy2008, Kondov2011, Jendrzejewski2012, Robert-de-Saint-Vincent2010, Krinner2012, Beeler2012,pasienski2010disordered,PhysRevLett.114.083002}.  A notable exception is disorder generated by impurity atoms \cite{Gadway2011}.  Optical speckle is produced by focusing a laser beam that has passed through a diffuser.  The atoms experience a potential energy shift proportional to the optical intensity of this light.  Speckle disorder has an exponential distribution of potential energies and a spatial autocorrelation that is approximately Gaussian. While in three dimensions the percolation threshold for a blue-detuned speckle field is negligible \cite{O'Holleran2008,pilati2010dilute}, potential minima appear in one and two dimensions that can classically trap the atoms. As a result, the percolation characteristics of a speckle potential become a significant constraint on attempts to observe Anderson localization in reduced dimensions.

This limitation is in part due to scaling symmetries unique to speckle disorder. A distinctive feature of a speckle field it that it has only one adjustable length scale, the correlation length $\zeta$. This characteristic leads to a simple scaling symmetry of the field: any change in scale of the system is equivalent to a suitable change in $\zeta$. The percolation threshold is necessarily scale invariant, and as a result it must be independent of $\zeta$. Thus, the critical energy for a percolation transition in a speckle field is always at a fixed fraction of the average potential energy $\Delta$, which simulation shows to be roughly 52\% in two dimensions \cite{Weinrib1982, Pezze2011,pilati2010dilute}. This dependence complicates observing AL in two dimensions using ultracold atoms.  In infinitely sized two-dimensional system, infinitesimal disorder will localize atoms with any kinetic energy \cite{PhysRevLett.42.673}.  The localization length grows exponentially with the particle energy \cite{Lee1985}, and thus relatively strong disorder is required to localize atoms on experimentally accessible length scales.  This leads to classical trapping of a wide range of particle energies, which may be difficult to separate from AL.

In light of this high, fixed percolation threshold, an alternative form of disorder that does not cause classical trapping is desirable.  The point-like disorder we investigate consists of individual Gaussian potential barriers with randomly distributed locations $\vec{x}_i$:
\begin{equation}
V(\vec{x})=\sum_i V_0 e^{- \left|\vec{x}-\vec{x}_i\right|^2 / w^2},
\label{disorder}
\end{equation}
where $V_0$ is the peak disorder energy, $i$ indexes the individual Gaussian potentials, and $w/\sqrt{2}$ is the rms width of an individual barrier.  This type of disordered potential was chosen for its simplicity and straightforward experimental realization in a cold atom experiment using holographic techniques \cite{dufresne2001computer,Pasienski2007, Gaunt2012}. As shown in Fig. 1, a random array of blue-detuned focused Gaussian laser beams can produce potential barriers as in Eq. \ref{disorder}.  The advantage of point-like disorder over a speckle potential is that the freedom to tune the density $n$ of potential barriers introduces a second length scale $n^{-1/2}$, and varying the ratio of $w$ and the average distance between scattering sites $n^{-1/2}$ allows the percolation threshold to be tuned.

\begin{figure}[h!]
\includegraphics[width=0.5 \textwidth]{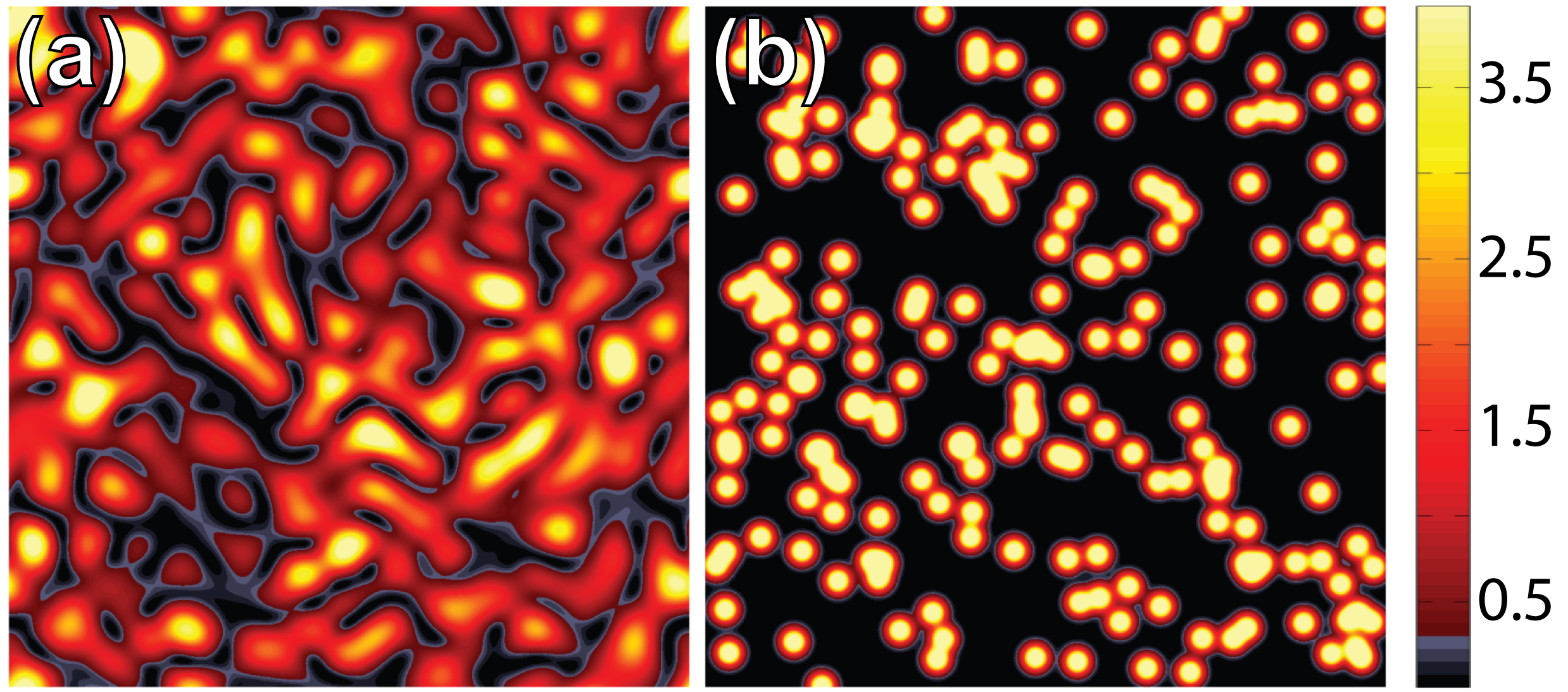}
\caption{Comparison of percolation in a speckle potential (a) and a sparse point-like disordered potential (b). The disorder potential energy is shown in false color.  The colorbar shows the potential energy in units of $\Delta$. Regions in grayscale correspond to energies less than 30\% the average disorder energy.  A classical particle with this energy would be trapped in a finite-sized region in the speckle potential, but is able to propagate indefinitely for the point-like disorder case.}
\label{fields}
\end{figure}

The impact of this tunability is shown visually in Fig.~\ref{fields}.  Potential landscapes are displayed for speckle and point-like disorder with the same average potential energy.  While the percolation threshold for this finite-sized realization of speckle disorder is $0.39\Delta$, the point-like disorder parameters were chosen to set the percolation threshold for the realization shown in Fig. 2b to approximately $0.06\Delta$.  Thus, particles with small kinetic energies compared with the disorder energy are free to propagate in the point-like disorder, while particles with relatively high kinetic energies are classically trapped by the speckle disorder.

\begin{figure}[h!]
\includegraphics[width=.45 \textwidth]{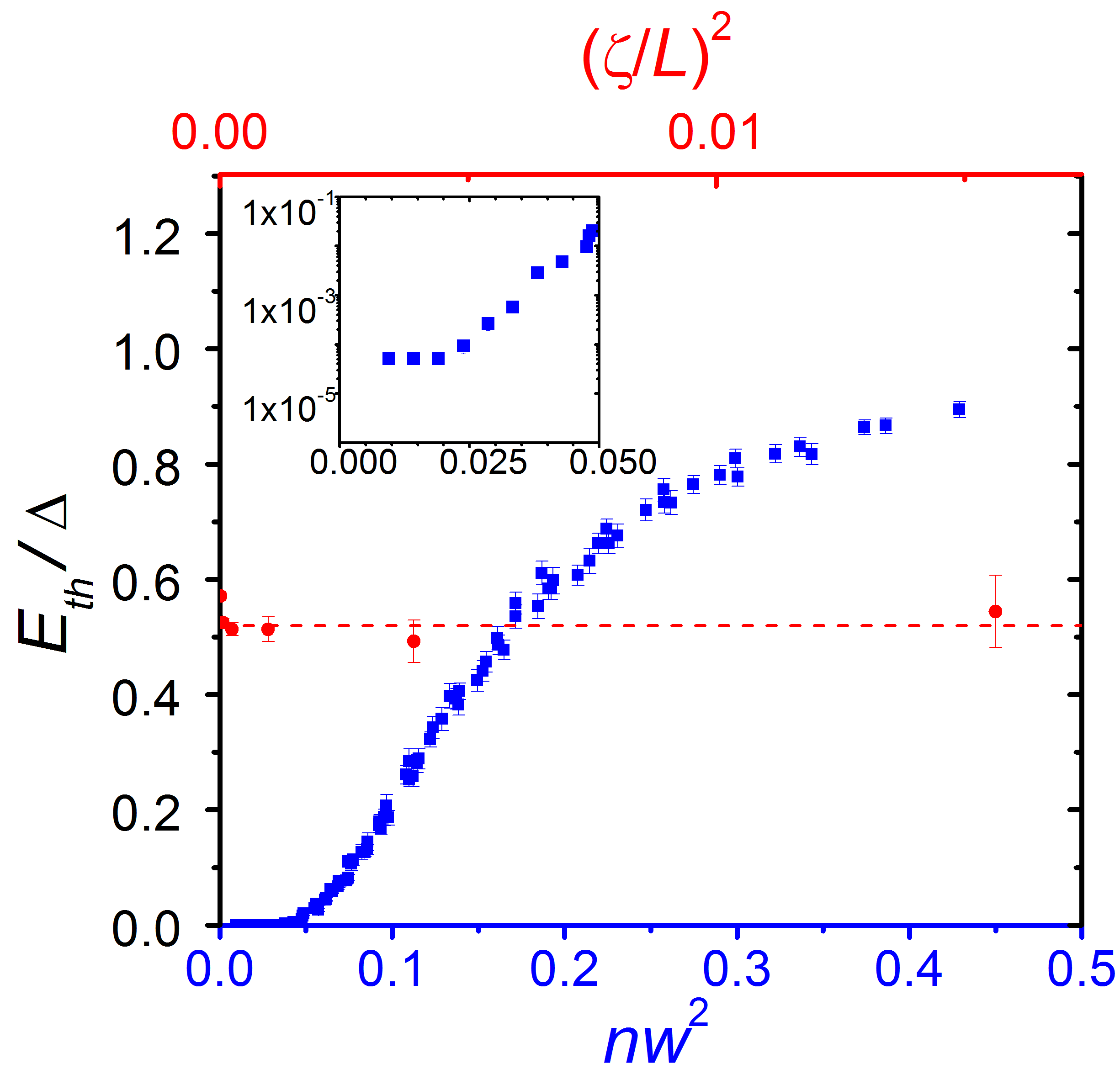}
\caption{Percolation threshold for point-like (blue squares) and speckle (red circles) disorder potentials. The dashed line is the known percolation threshold for speckle disorder, and the points are results of our simulation. For point-like disorder, $n$ and $w$ are independently varied, and for speckle disorder the correlation length was changed. The inset shows the onset of significant percolation for point-like disorder. The minimum detectable threshold level is $5\times10^{-5}$ in our simulation. The error bars show the standard error of the mean for the average taken over 50 disorder realizations at each point-like disorder point and 8 realizations for each speckle disorder case.}
\label{perc}
\end{figure}

We used a standard technique to calculate the dependance of the percolation threshold on the point-like disorder parameters.  Disorder potentials were numerically simulated, and the percolation threshold was determined by detecting the formation of a connected region spanning the simulation space by points below a threshold potential energy.  The percolation threshold $E_{th}$ was determined by averaging over independent realizations of the disorder potential.  While only approximating the percolation threshold, which is defined in limit of an infinite system, this method gives excellent agreement with the known value for speckle disorder [Fig. \ref{perc}].  In our simulation, a fixed system size $L$ is used and the number of Gaussian potential barriers and $w$ were independently varied.

As shown in Fig. \ref{perc}, the percolation threshold was found to be a function of only the dimensionless combination $nw^2$, as expected from simple space-filling considerations \cite{Mertens2012}.  For sufficiently dilute disorder (i.e., low $nw^2$), the percolation threshold is arbitrarily close to zero, and it monotonically increases with $nw^2$ towards the limit of the average disorder energy. Our interest is in the regime where the percolation threshold is lower than that of a speckle potential and ideally negligible, while maintaining sufficient density of scattering sites such that the particles are scattered many times within the system size.

We choose to concentrate on $nw^2\leq0.03$, which fulfills both criteria.  Under this condition, the percolation threshold is less than $2.5\times10^{-4} \Delta$, which is smaller than the percolation threshold for 3D speckle \cite{pilati2010dilute}.  The system size and $w$ are limited by experimental constraints such as optical power, imaging signal-to-noise ratio, and numerical aperture.  For the rest of this paper, we choose to use an experimentally feasible disorder size $w=400$~nm.  When discussing proposed experiments, we assume that the disorder potential and the light used to create it are limited to a $100\times100$~$\mu$m$^2$ area in order to estimate the requisite laser power.  We use $n=0.2$~$\mu$m$^{-2}$ (corresponding to $nw^2=0.03$) for comparing the localization lengths predicted by scaling theory using either the first Born approximation or the exact scattering differential cross section and for predicting a thermally averaged density profile.  In these cases, we choose the highest possible $n$ that avoids significant classical trapping in order to make AL effects as robust as possible.  For comparing the results of scaling theory and a simulation of the time-dependent Schr\"{o}dinger equation, we use $n=0.08$~$\mu$m$^{-2}$ (corresponding to $nw^2=0.013$), which is the largest $n$ compatible with our computational resources.  For $n=0.2$~$\mu$m$^{-2}$ and $n=0.08$~$\mu$m$^{-2}$, there are 2000 and 800 disorder peaks within the proposed $100\times100$~$\mu$m$^2$ area, respectively.

While we have shown that point-like disorder can avoid classical trapping, it must also lead to observable localization lengths to be a viable experimental option. In principle, atoms in an infinite two-dimensional system are localized by infinitesimal disorder \cite{Lee1985}, but only localization lengths smaller than the system size are physically meaningful and observable. Therefore, the rest of this paper is concerned with estimation of the localization length for point-like disorder and its dependance on the disorder properties and particle energy.

\section{Localization lengths: the Born approximation}

In this section we determine analytic expressions for the localization length in the limit where the scattered wavefunction amplitude is small.  Although we will show that this approach is inaccurate in the regime of interest, it is useful for developing physical insight.  Analysis of point-like disorder is particularly simple in this regime, because the differential cross section for scattering from a single disorder barrier can be determined using the first-order Born approximation. This differential cross section can be used to determine transport properties in a potential consisting of many Gaussian potentials, provided that there little spatial overlap.  Our procedure has three steps: we first find the Born approximation for the differential scattering cross section $d\sigma/d\theta$ of a single potential barrier. We use $d\sigma/d\theta$ to determine the Boltzmann transport mean free path $l_B$, which characterizes the diffusive properties for particle transport in the disordered potential. Finally, we use scaling theory to estimate the localization length $\xi$ from $l_B$.

In two dimensions, we write the (unnormalized) scattered wavefunction $\psi(\vec{x})$ as
\begin{equation}
\psi(\vec{x})=e^{i\vec{k}\cdot\vec{x}}+f\left(\theta\right)\frac{e^{ikr}}{\sqrt{r}},
\label{scatter}
\end{equation}
where $\vec{r}$ is the radial coordinate, $\vec{k}$ is the incoming wavevector, $f\left(\theta\right)$ is the scattering amplitude, and $\theta$ is the scattering angle. We assume a free-particle dispersion so that $k=\sqrt{2m\epsilon_k}/\hbar$, where $m$ is the mass of the particle, $\epsilon_k$ is the kinetic energy of the atom, and $h=2\pi\hbar$ is Planck's constant. Given the scattering amplitude in the Born approximation
\begin{equation}
f\left(\theta\right)=-\frac{m e^{i\frac{\pi}{4}}}{\hbar^2\sqrt{2\pi k}} \int e^{i\left(k\hat{r}-\vec{k}\right)\cdot\vec{x}} V\left(\vec{x}\right) d^2\vec{x}
\end{equation}
($\hat{r}$ is a unit vector that points in the scattered direction), the differential cross section for a single Gaussian potential $V(\vec{x})=V_0 e^{-r^2/w^2}$ is
\begin{eqnarray}
\frac{d\sigma}{d\theta}&=&\left|f\left(\theta\right)\right|^2 \nonumber\\
&=&\frac{\pi}{8k}\left( \frac{2m}{\hbar^2} \right) ^2 w^4 V_0^2 e^{-2w^2 k^2\sin^2\frac{\theta}{2}},
\label{diffxs}
\end{eqnarray}
where $\theta$ is the scattering angle.  We calculate the Boltzmann mean free path $l_B$, the distance over which the direction of momentum remains correlated \cite{Kuhn2007}, using Eq.~\ref{diffxs} for the differential cross section and the density of scattering sites $n$:
\begin{equation}
l_B=l_s \left[ 1-\int_0^{2\pi}\cos{\theta}\left( \frac{1}{\sigma} \frac{d\sigma}{d\theta}\right)d\theta\right] ^{-1},
\label{lB}
\end{equation}
where the elastic mean free path is $l_s=(n\sigma)^{-1}$, and the total cross section is $\sigma=\int_0^{2\pi}\left(d\sigma/d\theta\right)d\theta$. Combining equations \ref{diffxs} and \ref{lB}, we find
\begin{equation}
l_B=\frac{1}{\sqrt{2\pi^3}}\frac{1}{nw} \left( \frac{\hbar^2 k^2}{2m}\frac{1}{V_0}  \right)^2
\label{Boltzmann}
\end{equation}
for the Boltzmann mean free path.

The connection between $l_B$ and the localization length $\xi$  is given in the $(kl_B)^{-1} \ll 1$, $L\rightarrow \infty$ limit by scaling theory \cite{Lee1985} as
\begin{equation}
\xi=l_Be^{\frac{\pi}{2}kl_B}.
\label{xi}
\end{equation}
Thus, given the properties of the disorder and atoms, we may calculate a localization length valid in these limits using the Born approximation.  We choose to focus on an experimentally realistic situation. In an experiment in which the disorder is generated by the dipole force from a far-detuned laser, limited laser power constrains the maximum average potential energy, rather than the value of $V_0$ for individual potential barriers. Therefore, we investigate the case in which $\Delta$ is fixed, but the  density of disorder sites $n$ and $V_0$ vary inversely.

\begin{figure}[htb]
\begin{center}
\includegraphics[width=0.5 \textwidth]{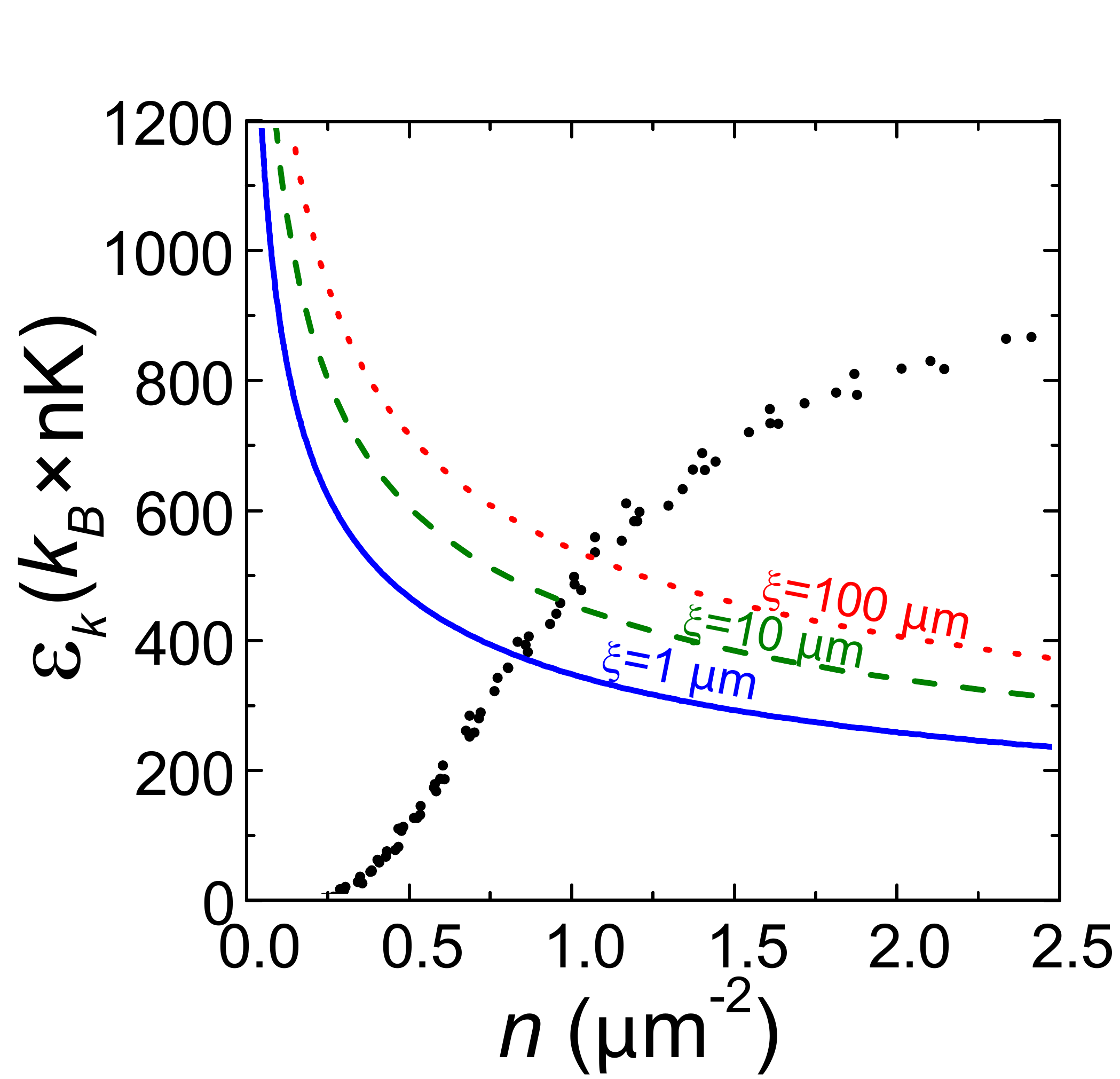}
\caption{Localization lengths according to scaling theory within the Born approximation.  Contour lines for three localization lengths are displayed using the experimental parameters described in the main text.  The percolation threshold in terms of $\epsilon_k$ is shown using solid circles.  If $\epsilon_k>E_{th}$, then AL will not be observable and the atoms will be classically trapped.  Because the average disorder potential energy is fixed at $k_B\times1000$~nK, $V_0$ varies according to $V_0=k_B\times\left(1989/n[\mu\mathrm{m}^{-2}]\right)$~nK.}
\label{wslengths}
\end{center}
\end{figure}

To get a concrete sense of these predictions, for disorder generated by a 2~W, 532~nm laser focused to a Gaussian envelope with a 170~$\mu$m waist (which has been employed in experiments on 3D AL \cite{Kondov2011}), realistic disorder parameters are $w=400$~nm and average disorder strength $\Delta=k_B\times1000$~nK.  The peak potential $V_0$ is determined by the relation $V_0=\Delta/\pi n w^2$. The resulting localization lengths, as a function of particle energy $\epsilon_k$ and $n$, are shown in Fig. \ref{wslengths} along with the percolation threshold. For the contour lines shown in Fig.~\ref{wslengths}, the value of $kl_B$ varies from 0.05--0.3.  A localization length of 100~$\mu$m, which would result in high imaging signal-to-noise ratio and minimal effects from the disorder envelope, can be well-separated from classical trapping when there is less than approximately one potential barrier per square micron.  Furthermore, a regime exists for $n<0.25\mu$m$^{-2}$ in which classical trapping is irrelevant and small localization lengths exist for experimentally accessible particle energies.

While this standard approach to determining localization lengths suggests that two-dimensional localization may be observable using ultracold atoms and point-like disorder, it is unclear that the Born approximation will be valid in an experiment.  Given the parameters explored in Fig.~\ref{wslengths}, accessible localization lengths necessarily involve $V_0 > \epsilon_k$, and thus the Born approximation is suspect.  The precise limits of the Born approximation are not always obvious \cite{Gottfri}.  Unlike the three-dimensional case, the Born approximation is never valid in two dimensions as $k \rightarrow 0$. In two dimensions and for $kw\ll1$, a simple analysis suggests that the Born approximation may be valid for
\begin{equation}
g= \frac{4mw}{\hbar^2}\frac{V_0}{k} \ll 1.
\end{equation}
The smallest value for $g\approx7.5$ occurs in the upper right hand corner of Fig.~\ref{wslengths}, and thus the Born approximation is not satisfied for this range of parameters.  Similar considerations hold for the high energy (i.e., $kw\gg1$) regime.  Thus, while the Born approximation can supply physical insight, the localization lengths computed in this section may not be accurate.

Using the Born approximation will generally underestimate the localization length.  The differential cross section in the Born approximation [Eq. \eqref{diffxs}] depends quadratically on the strength $V_0$ of a scattering site, which accounts for the sharp decrease in localization length in Fig. \ref{wslengths} near $n=0$, where the disorder potential is concentrated in few, very strongly scattering sites. However, once $V_0$ is much greater than the particle kinetic energy, further increasing $V_0$ must have diminishing effects on scattering.  Underestimating the localization length is a critical problem because large localization lengths may not be observable in an experiment.   We determine the precise deviation from the Born approximation behavior using numerical simulation.  As we will show in the next section, the Born approximation and scaling theory as applied here can underestimate the localization length by more than three orders of magnitude in the regime we propose to explore experimentally.

\section{Beyond the Born Approximation}

In this section we implement numerical simulations of scattering and localization, and use them to determine improved estimates of localization lengths. We rely on two complementary approaches. The first, simplest approach is a single scattering simulation: we numerically determine the exact differential cross-section for a single scattering event, and then use this in place of the analytic result (Eq.~\ref{diffxs}) to compute $l_B$ and $\xi$ as in Sec. III. While the values obtained using this method remain valid only for $kl_B\gg1$, they have two important advantages over the analytic solutions: in practice the error is much less severe at low energy, and, critically, the assumptions that entered into the scaling theory derivation are such that it provides an upper bound on the localization length \cite{Lee1985}.

\begin{figure}[h!]
\begin{center}
\includegraphics[width=0.5 \textwidth]{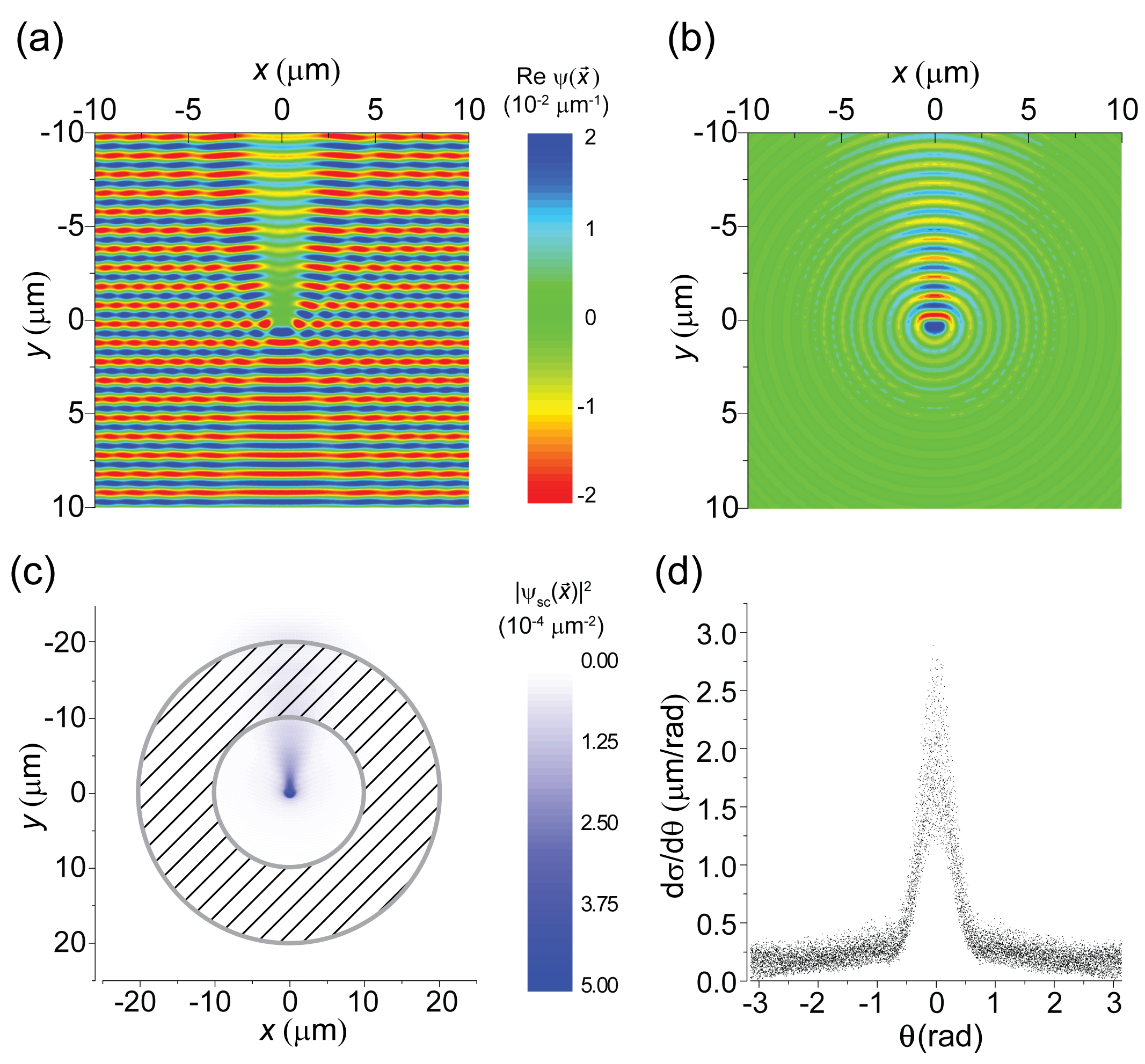}
\caption{Process of extracting scattering properties from a simulating propagation of a plane wave for the time-dependent Schr\"{o}dinger equation. The initial wavevector $\vec{k}$ is in the $y$ direction.  (a) Real part of the wavefunction $\psi(\vec{x},\tau)$ shown in false color.  For all of the data shown in this figure, $\epsilon_k=k_B\times240$~nK, $V_0=k_B\times1000$~nK, and $w=400$~nm.  Violation of the Born approximation is evident as significant distortion of the initial plane wave by the potential barrier at the origin (not pictured).  (b) Real part of the scattered wavefunction $\psi_{sc}(\vec{x})$ obtained by subtracting a plane wave propagated in free space from the data shown in (a).  The color bar applies to the simulated wavefunctions shown in (a) and (b).  (c) The scattered probability density $\left|\psi_{sc}(\vec{x})\right|^2$ is shown as a density plot. The cross-hatched region is sampled to determine that differential cross section. (d) Differential cross section obtained from the data in (c).  Points are shown at fixed $\theta$ and different $r$ within the circular annulus.}
\label{sssim}
\end{center}
\end{figure}

To apply this method, we numerically extract $d\sigma/d\theta$ by a straightforward application of scattering theory (Fig.~\ref{sssim}). We first simulate, via the split-step Fourier method \cite{Xu2003}, a plane wave scattering from a single potential barrier (see Appendix A).  A wavefunction $\psi(\vec{x},t=0)=e^{iky}/L$ with energy $\epsilon_k=\hbar^2k^2/2m$ and wavevector $k\hat{y}$ is prepared at initial time $t=0$ in the simulation space, which has sides of length $L$.  The wavefunction is propagated forward in timesteps $\delta t$ according to
\begin{eqnarray}
\psi(\vec{x}&,&t+\delta t)= e^{-i \frac{\delta t}{2} V(\vec{x},t)}\cdot\nonumber\\
& & \mathcal{F}^{-1}\left \{ e^{-i \delta t \frac{\hbar^2 k^2}{2m} }\mathcal{F}\left \{ e^{-i \frac{\delta t}{2} V(\vec{x},t)} \psi (\vec{x},t) \right \} \right \},
\label{splitstep}
\end{eqnarray}
where $\mathcal{F}$ represents a Fourier transform. The timestep $\delta t$ is chosen to minimize numerical instability and such that $\delta t \cdot V/\hbar\ll1$ and $\delta t \cdot \hbar k^2 /2m\ll1$.

While propagating the wavefunction forward in time, the Gaussian potential $V(\vec{x},t)=V_0 e^{-r^2/w^2} \left(1-e^{-t/0.2\tau}\right)$ is slowly ramped on. We determined that the time over which the barrier was turned on did not significantly change the simulation results.  After a total propagation time from $\tau=1$--7~ms (depending on the group velocity), a wavefunction distorted by scattering from the potential barrier is produced, as shown in Fig.~\ref{sssim}a.  For the $V_0$ and $\epsilon_k$ chosen in Fig.~\ref{sssim}, the simulated wave is highly distorted, signaling a violation of the Born approximation.

The scattered wave $\psi_{sc}=\psi(\vec{x},t)-\psi'(\vec{x},t)$ is recovered using the unscattered wave $\psi'(\vec{x},t)$, which is determined by repeating the simulation with $V(\vec{x},t)=0$.  The scattering amplitude $f_r(\theta)$ at radius $r$ is computing according to Eq.~\ref{scatter} as $f_r(\theta)=e^{-ikr}\sqrt{r}\psi_{sc}(\vec{x})$.  As shown in Fig.~\ref{sssim}, $f_r(\theta)$ is sampled in a circular annulus centered on the origin.  The outer and inner radii of this annulus were set to minimize boundary effects and to achieve the asymptotic scattering regime.  The differential cross section $d\sigma/d\theta$---calculated by averaging $\left|f_r(\theta)\right|^2$ across all $r$ within the annulus at fixed $\theta$---is numerically integrated to produce the cross section $\sigma$.

\begin{figure}[h!]
\begin{center}
\includegraphics[width=.4 \textwidth]{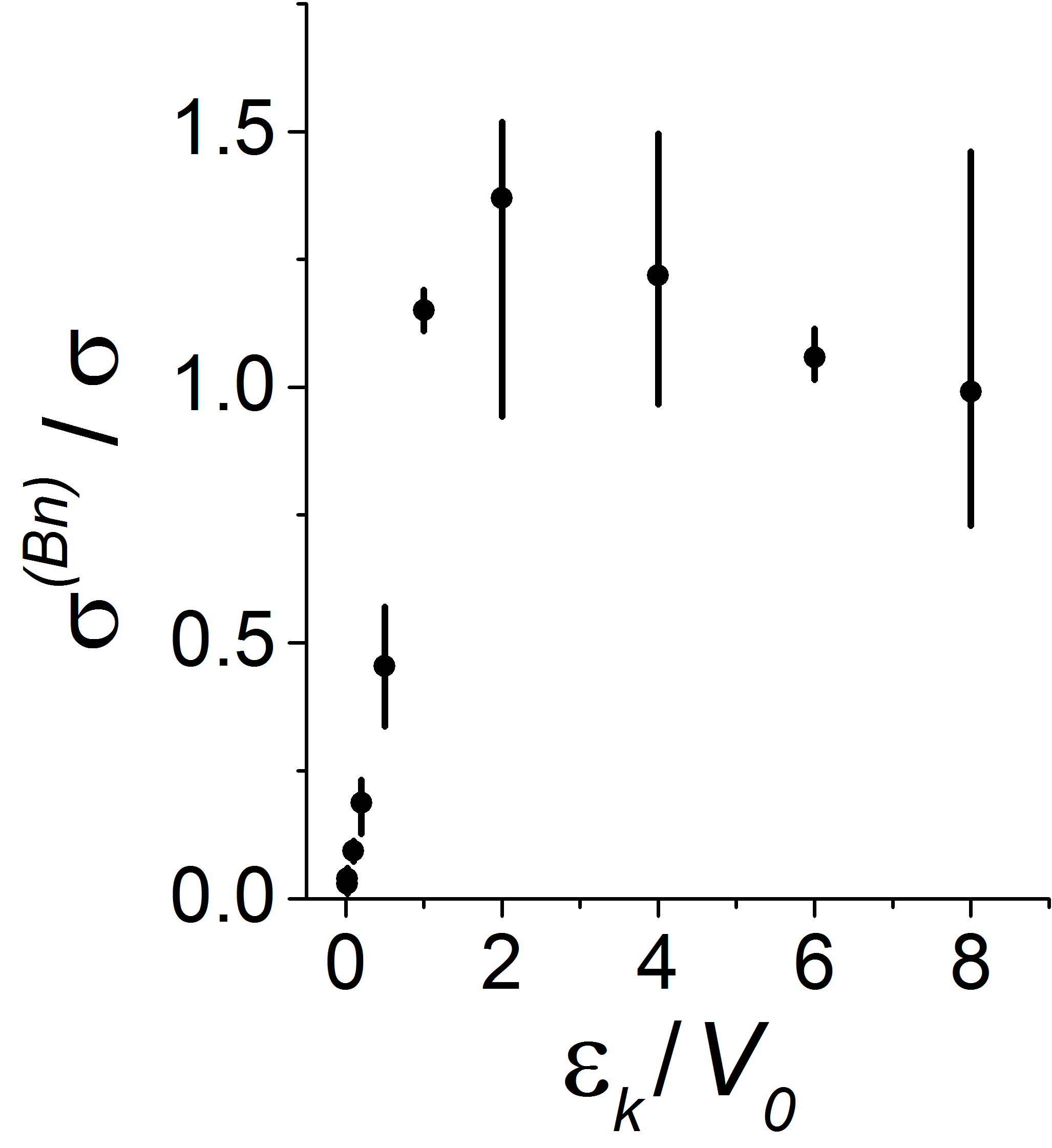}
\caption {Ratio of total collision cross section determined from the Born approximation $\sigma^{(Bn)}$ to $\sigma$ calculated using the exact differential cross section for $w=400$~nm and $V_0=k_B\times50$~nK. The error bars show the spread in values determined across the inner and outer radii of the annulus displayed in Fig. 5c and include the effects of numerical errors and deviation from the far-field limit.} 
\label{simtest1}
\end{center}
\end{figure}

A comparison between the total cross section $\sigma$ determined using the exact $d\sigma/d\theta$ and using $d\sigma/d\theta$ from the Born approximation is shown in Fig.~\ref{simtest1}.  The two approaches agree at high kinetic energy.  At low energies, however, the Born approximation fails and underestimates the cross section.

\begin{figure}[h!]
\begin{center}
\includegraphics[width=0.4 \textwidth]{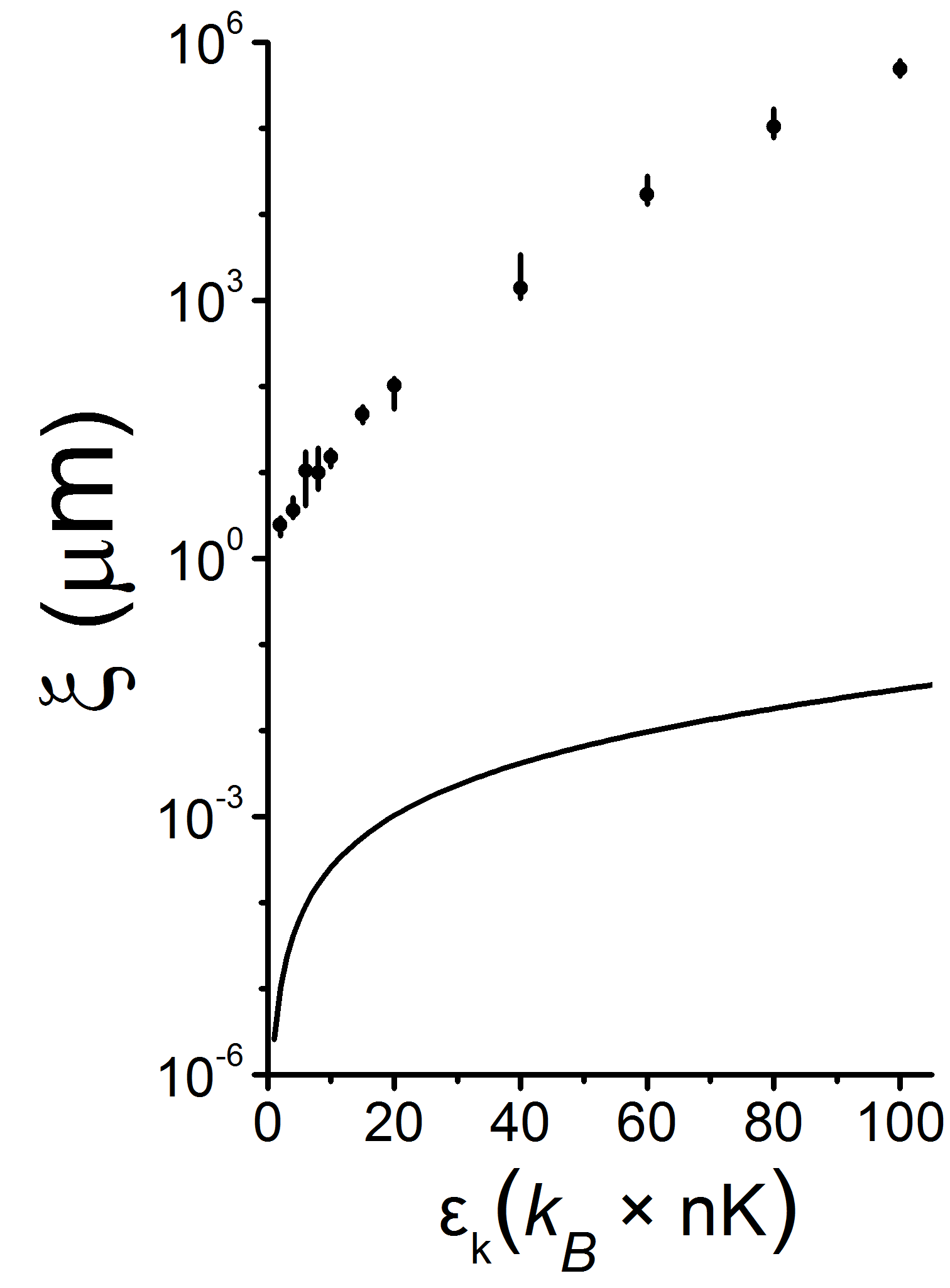}
\caption{Localization lengths predicted using scaling theory and the exact differential cross section (circles) and the Born approximation (solid line).  The density for the disorder potential is $n=0.2$~$\mu$m$^{-2}$, the average disorder potential energy $\Delta=k_B\times1000$~nK, and the Gaussian barrier height $V_0=k_B\times9950$~nK. The error bars are determined using the spread in the differential cross section determined across the inner and outer radii of the annulus displayed in Fig. 5c.}
\label{explengths}
\end{center}
\end{figure}

An improved scaling theory prediction for the localization length $\xi$ using the exact differential cross section is shown in Fig.~\ref{explengths}.  We fix $n=0.2$~$\mu$m$^{-2}$, which corresponds to a vertical slice in Fig.~\ref{wslengths}.  The localization length predicted using the exact differential cross section exceeds the Born-approximation prediction by more than three orders of magnitudes at all particle energies.  Furthermore, the localization length is always larger than the characteristic length scales of the disorder, while the Born approximation predicts unphysical localization lengths much smaller than $w$ and $1/\sqrt{n}\approx$~2~$\mu$m.  For the disorder parameters explored here and a 100~$\mu$m system size, localization with be visible for point-like disorder for particles with energies less than $k_B\times20$~nK.  While for a speckle potential with the same $\Delta$ the percolation threshold is $k_B\times500$~nK, the percolation threshold for this point-like disorder potential is approximately $k_B\times0.25$~nK, and thus AL will be the dominant influence on transport.

\begin{figure}[h!]
\includegraphics[width=0.45 \textwidth]{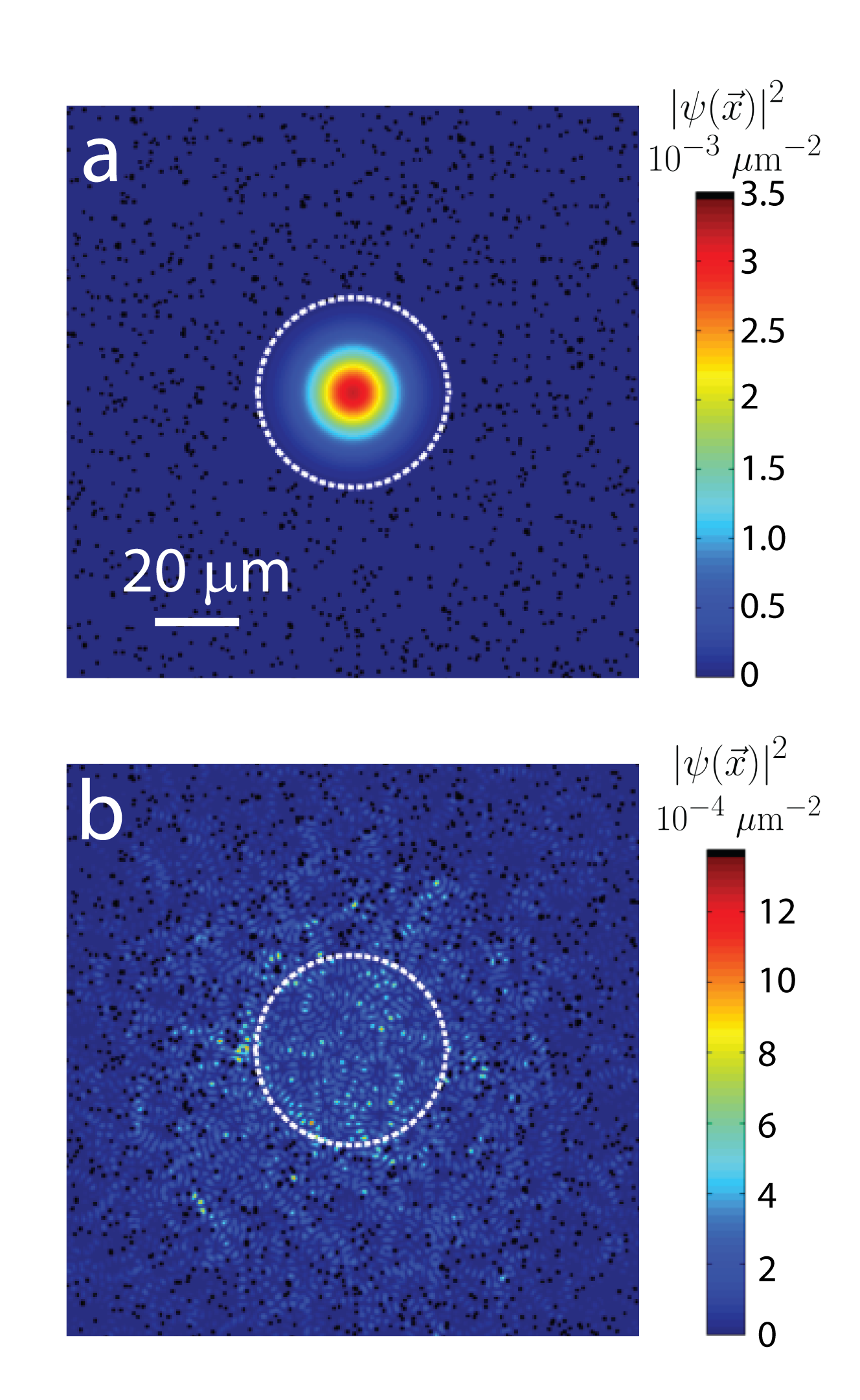}
\caption{Simulated probability density distributions shown in false color from a time-dependent simulation of localization.   A Gaussian wavepacket initialized at $t=0$ in the disorder-free region marked by the white dashed line is shown in (a).  The wavefunction propagated forward in time for 146~ms is shown in (b).   The potential barriers that constitute the disorder potential are magnified and marked in black for clarity.  For these simulations, $\epsilon_k=k_B\times25$~nK, $V_0=k_B\times1000$~nK, $w=400$~nm, and $n=0.08\,\mu$m$^{-2}$.}
\label{fullsim}
\end{figure}

As a second approach to determining localization lengths, we performed complete time-dependent simulations of a wavepacket propagating through a point-like disorder potential (see Appendix B).  This approach is inspired by the method used to observe AL in 1D \cite{Billy2008, Roati2008} and 3D \cite{Kondov2011, Jendrzejewski2012,PhysRevLett.111.145303} ultracold gases: atoms initially confined in a small region of space by a trap are allowed to expand into a disordered potential.  The atoms propagate through the disorder potential, and, through scattering, eventually adopt a localized profile.  We numerically simulate independent realizations of a potential according to Eq.~\ref{disorder}, with a disorder-free, circular region of radius $R$ centered on the origin (see Fig.~\ref{fullsim}a).  As an initial condition, a Gaussian wavepacket $\psi(\vec{x},t=0)\propto\exp\left(ikr-r^2/2(R/2.2)^2\right)$ with $kR\gg1$ is prepared in the disorder-free region. The wavenumber $k$ is varied from 0.07--0.4~$\mu$m$^{-1}$, which corresponds to a kinetic energy from $k_B\times$~3--100~nK.  The shift in the average kinetic energy of the wavepacket resulting from the gaussian envelope is less than 2\% of $\epsilon_k$.

\begin{figure}[h!]
\includegraphics[width=0.5 \textwidth]{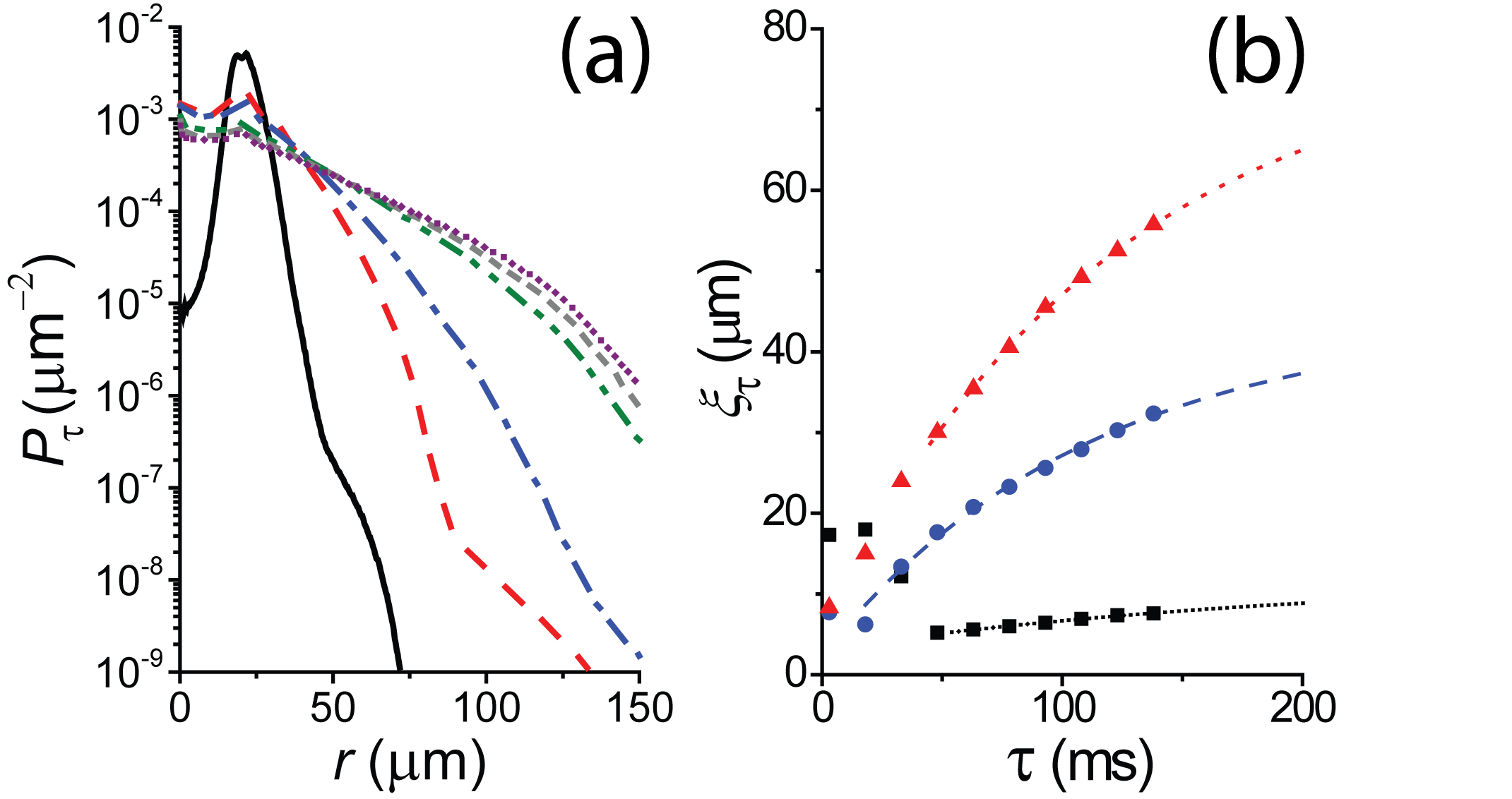}
\caption{Procedure to determine $\xi$ from the time-dependent simulations of localization. (a) Sample radial probability density showing the approach to a steady-state profile. The propagation times are $\tau=3$~ms (solid black line), 15~ms (red dashed line), 30~ms (blue dotted-dashed line), 90~ms (green dashed-dotted-dotted) line, 120~ms (short dashed line), and 150~ms (purple dotted line). For the data shown in this figure, $\epsilon_k=k_B\times100$~nK, $V_0=k_B\times1000$nK, $w=400$~nm, and $n=0.08$~$\mu$m$^{-2}$.  (b) Fitted exponential decay lengths $\xi_\tau$ from data such as those in (a) for $\epsilon_k=k_B\times100$~nK (red triangles), 35~nK (blue circles), and 3~nK (black squares).  Only a fifth of the points generated for each value of $\epsilon_k$ are shown.  The uncertainty in the points is too small to be visible.   Fits of these data to an exponential function (lines) are used to extract the asymptotic value of the localization length $\xi$.}
\label{fullsim2}
\end{figure}

The wavefunction is propagated according to Eq.~\ref{splitstep} for total time $\tau=146$~ms using timesteps $\delta t=14.6\mu$s in order to resolve the effects of the disorder potential energy on the wavefunction.  During propagation, the wavepacket disperses into the disorder potential, eventually forming a localized state (Fig.~\ref{fullsim}b).  We perform an angular average of $\left|\psi(\vec{x},\tau)\right|^2$ at fixed radius $r$, which is then averaged over multiple realizations of the disorder potential to produce the disorder-averaged radial probability density $P_\tau(r)$.  Sample $P_\tau(r)$ for fixed disorder parameters $V_0=k_B\times1000$~nK, $w=400$~nm, and $n=0.08$~$\mu$m$^{-2}$ and kinetic energy $\epsilon_k=k_B\times 100$~nK are shown in Fig.~\ref{fullsim2}a. An exponential distribution at large $r$ is rapidly achieved, which then expands and relaxes at long times to a static, localized state.  We fit $P_\tau(r)$ to an exponential decay for $r=43$--50~$\mu$m for $\epsilon_k\leq k_B\times$10~nK and $r=57$--68~$\mu$m for $\epsilon_k>k_B\times$10~nK at each $\tau$ in order to determine the characteristic size $\xi_\tau$ of the propagated wavefunction.  The asymptotic localization length $\xi$ (that would be observed in an experiment probing atoms expanding through a disordered potential) is determined by fitting $\xi_\tau$ to an exponential function, discarding points at short times.

\begin{figure}[h!]
\includegraphics[width=.5 \textwidth]{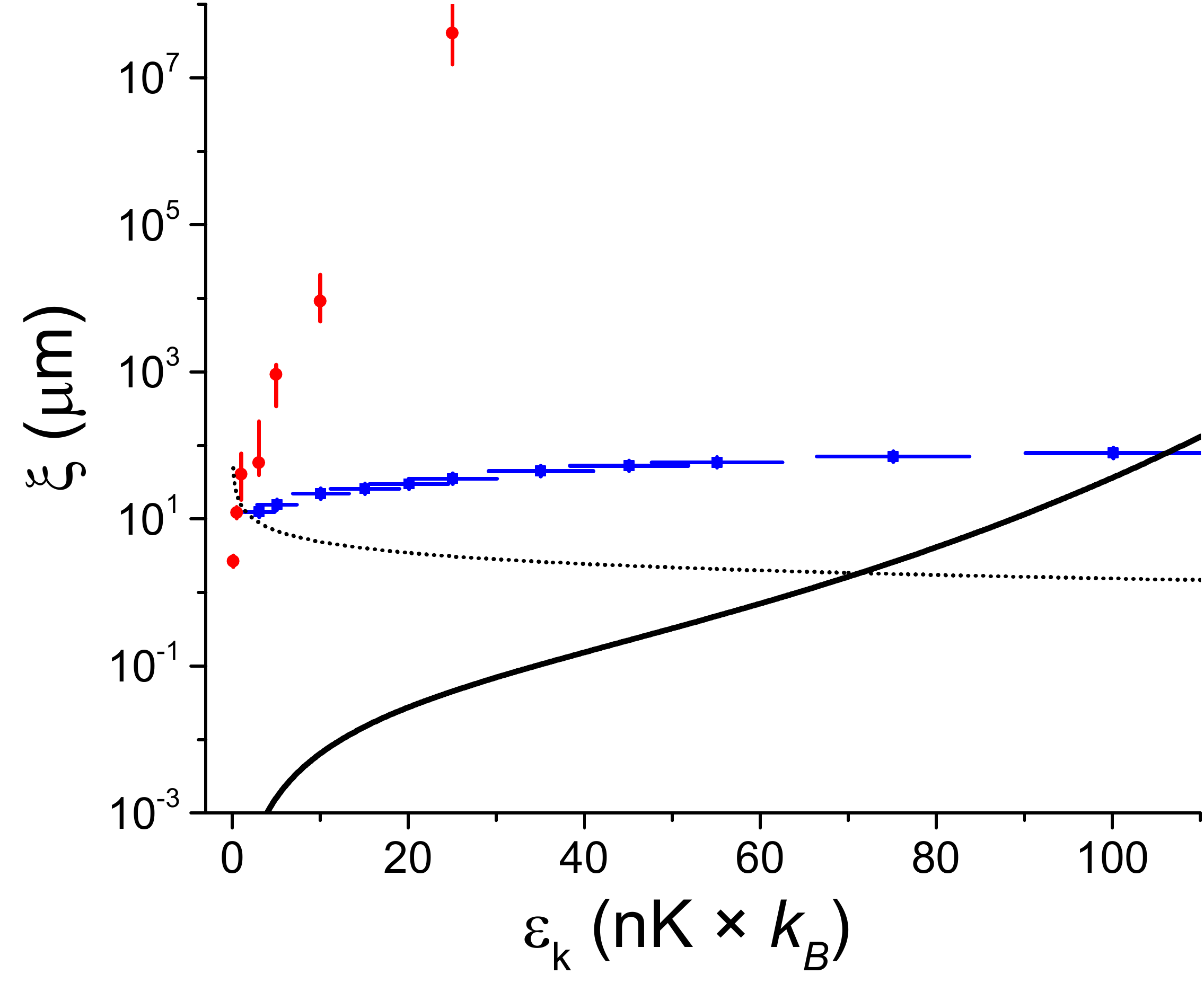}
\caption{Comparison of localization lengths from a numerical simulation of the time-dependent Schr\"{o}dinger equation (blue squares), from the perturbative correction to the independent scattering transport properties (red circles), and from the Born approximation (black line). For comparison, the de Broglie wavelength $2\pi/k$ is shown as a dotted line.  The parameters used for this plot are $V_0=k_B\times1000$~nK, $w=400$~nm, and $n=0.08\,\mu$m$^{-2}$. The error bars for the circles shown the impact of the difference in differential cross section across the inner and outer radii for the annulus shown in Fig. 5c. and the standard deviation in the energy (resulting from the Gaussian envelope of the wavepacket) for the squares.}
\label{simcompare}
\end{figure}

These numerical simulations enable us to benchmark the individual scattering technique against a calculation without any inherent approximations. A comparison is shown in Fig. \ref{simcompare}.  At the lowest energies, the single scattering approximation predicts localization lengths smaller than the de Broglie wavelength $2\pi/k$ of the particle, which occurs because $n\sigma<2\pi/k$ and the approximation that $l_s=\left(n\sigma\right)^{-1}$ is violated.  In contrast, the results of the time-dependent simulation always produce localization lengths greater than the de Broglie wavelength.  As expected, the single scattering approximation provides an upper bound on the numerically determined localization length. At low energies, the single scattering approximation and results of the time-dependent simulations agree within an order of magnitude.  The Born approximation, however, is four orders of magnitude smaller than the numerically simulated $\xi$ at low energies.

\begin{figure}[h!]
\includegraphics[width=.5 \textwidth]{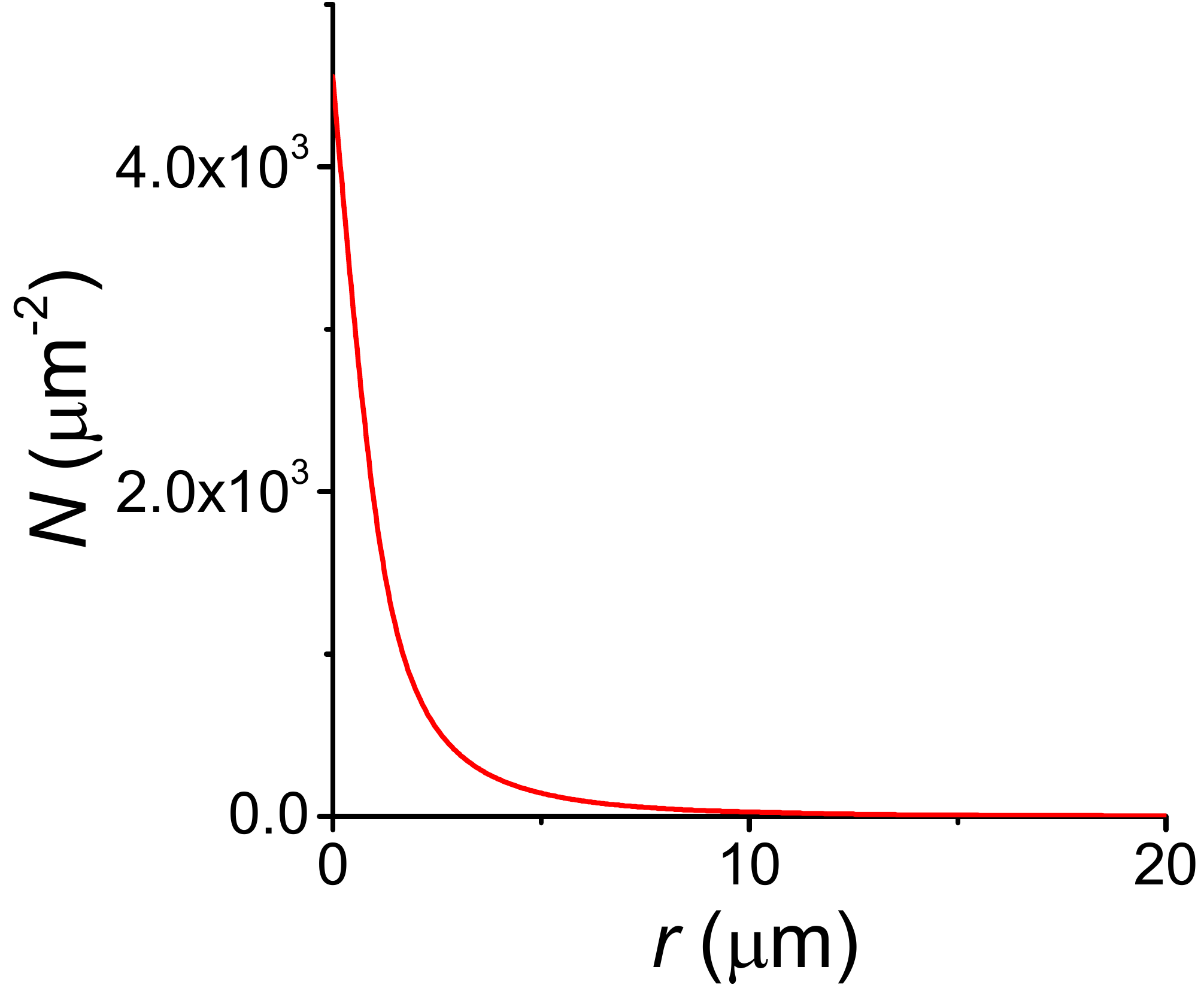}
\caption{Thermal density profile for $V_0=k_B\times9947$~nK, $w=400$~nm, $n=0.2$~$mu$m$^{-2}$, and $T=10$~nK.  Here, $N$ is the two-dimensional number density for 10,000 particles.}
\label{thermal}
\end{figure}

Finally, we calculate a localized density profile for a thermal gas of particles under conservative conditions (i.e., $n=0.2\,\mu$m$^{-2}$) using the simulated localization lengths from Fig. \ref{explengths}, corresponding to the single scattering approximation and exact scattering cross section.  We assume that the density profile associated with each energy in the thermal ensemble is a radially symmetric exponential function centered at the origin with a decay length determined by a fit to the data in Fig. \ref{explengths}. The average density profile of a localized gas of 10,000 Maxwell-Boltzmann particles at temperature $T=10$~nK is shown in Fig. \ref{thermal}. Dimensionality plays a helpful role, since, unlike in three dimensions, the lowest particle energies (with the smallest localization lengths) are the most probable. Hence, most of the particles localize within a relatively small, experimentally accessible area.  For example, at $T=10$~nK, 90\% of the particles are localized within a radius of 20~$\mu$m.  Furthermore, just 2.5\% of the particles have energies below the percolation threshold and are classically trapped.  A gas released into the disorder potential with a time-independent profile similar to that shown in Fig.~\ref{thermal} is straightforward to image with high signal-to-noise ratio and would provide a clear signature of two-dimensional AL. Since the single scattering approximation is an upper bound on the localization length observed in an expansion experiment, the localized thermal density profile should be smaller in extent than what is shown in Fig.~\ref{thermal}.

\section{Conclusion}

We have identified an experimentally feasible approach to observation two-dimensional AL using ultracold atoms using point-like disorder.  Our approach requires nanoKelvin-scale temperatures in conjunction with a dilute, strongly scattering disordered potential.  Using a single scattering approximation and exact numerical simulations, we predict observable localization lengths for realistic particle energies.  This method can be extended to interacting systems such as a gas composed of two spin states of fermionic atoms, which may enable exploration of the analog of two-dimensional metal-insulator transitions \cite{kravchenko2010,goldman2010,RevModPhys.73.251}.

\appendix
\section{Individual Scattering Simulation}

In simulations of scattering from a single Gaussian potential barrier, square grids were employed ranging from $L^2=3600$ $\mu$m$^2$ to 10000 $\mu$m$^2$ with discretization sizes of 0.01 to 0.25~$\mu$m.  The size of the time steps was scaled with the shorter of the two physical length scales in the system, the disorder width $w$ and the particle de Broglie wavelength $2\pi/k$, according to  $\delta t$=50.1 ns/$\mu$m. Approximately 16,000 time steps were taken, which allowed the wavefunction initially at the center of the potential to scatter outward through roughly $L/2$, independent of the group velocity. Propagation for longer times resulted in significant errors induced by boundary and finite-size effects, manifest as a periodic modulation of the scattered wave. The scattered wavefunction was sampled between radii of 3--5~$\mu$m in order to determine the scattering amplitude.

\section{Time-Dependent Simulation of Anderson Localization}

We used a $1536\times1536$ discretization of the system area $349\times349$~$\mu$m$^2$ and populated the region $r>R$ with pointlike disorder given by
\begin{equation}	
	V({\vec{x}})=\sum_i V_0e^{-\left(\frac{\vec{x}-\vec{x}_i}{w}\right)^2},
\end{equation}
where the coordinates $\vec{x}_i$ are chosen randomly with the constraint that $r_i>R=21.8$~$\mu$m. We choose $w=400$ nm, $V_0=k_B\times1000$~nK, and $n=0.08$~$\mu$m$^{-2}$, which corresponds to disorder strong enough to produce localization within the simulation space while remaining computationally tractable.

We include an imaginary potential component near the edges of the system grid in order to absorb probability current approaching the periodic boundary.  This enables dynamical simulations to continue for the long times necessary to observe localization.  The absorbing boundary potential is given by
\begin{equation}
V=
\begin{cases}
-iB\frac{r}{r_{max}}, & |x|,|y| >129\, \mu \text{m} \\
0,&\text{otherwise}\\
\end{cases}
\end{equation}
with $r_{max}=174$~$\mu$m and $B=k_B \times \left(5\times 10^{-5}\right)$~nK.

\begin{acknowledgements}
We thank W. Shirley for assistance with numerical simulations of wavepacket propagation.  We acknowledge support from the NSF under grant PHY12-05548 and from the ARO under grant W9112-1-0462.
\end{acknowledgements}

\clearpage

\bibliographystyle{apsrev4-1}
\bibliography{library}

\end{document}